\documentclass[preprint]{aastex}

\slugcomment{ApJ accepted on 2013 September 17}

\usepackage{color}

\shorttitle{17P/Holmes: Contrast in activity between before and
after the 2007 outburst}
\shortauthors{Ishiguro et al.}

\usepackage{amstext}

\begin{document}

\title{17P/Holmes: Contrast in activity between before and after
the 2007 outburst}

\author{Masateru \textsc{Ishiguro}\altaffilmark{1}}
\affil{Department of Physics and Astronomy, Seoul National University,\\
Gwanak, Seoul 151-742, Republic of Korea}
\altaffiltext{1}{Visiting Scholar, Department of Earth and Space
Sciences, University of California at Los Angeles, 595 Charles Young
Drive East, Los Angeles, CA 90095-1567, USA}
\email{ishiguro@astro.snu.ac.kr}

\author{Yoonyoung \textsc{Kim}}
\affil{Department of Physics and Astronomy, Seoul National University,\\
Gwanak, Seoul 151-742, Republic of Korea}

\author{Junhan \textsc{Kim}}
\affil{Department of Astronomy \& Steward Observatory, University of
Arizona, 933 N Cherry Ave., Tucson, AZ 85721, USA}


\author{Fumihiko \textsc{Usui}}
\affil{Department of Astronomy, Graduate School of Science, 
The University of Tokyo,\\ 7-3-1 Hongo, Bunkyo-ku, Tokyo 113-0033, Japan}

\author{Jeremie J. \textsc{Vaubaillon}}
\affil{Observatoire de Paris, I.M.C.C.E., Denfert Rochereau, Bat. A.,
FR-75014 Paris, France}

\author{Daisuke \textsc{Ishihara}}
\affil{Department of Physics, School of Science, Nagoya University,
Furo-cho, Chikusa-ku, Nagoya, Aichi 464-8602, Japan}

\author{Hidekazu \textsc{Hanayama}}
\affil{Ishigakijima Astronomical Observatory, National Astronomical
Observatory of Japan, Ishigaki, Okinawa 907-0024, Japan}

\author{Yuki \textsc{Sarugaku}, Sunao \textsc{Hasegawa}}
\affil{Institute of Space and Astronautical Science, Japan Aerospace
Exploration Agency, \\ 3-1-1 Yoshinodai, Chuo-ku, Sagamihara,
Kanagawa 252-5210, Japan }

\author{Toshihiro \textsc{Kasuga}}
\affil{National Astronomical Observatory, 2-21-1 Osawa, Mitaka, Tokyo
181-8588, Japan}

\author{Dhanraj S. \textsc{Warjurkar}, Ji-Beom \textsc{Ham}}
\affil{Department of Physics and Astronomy, Seoul National University,\\
Gwanak, Seoul 151-742, Republic of Korea}

\author{Jeonghyun \textsc{Pyo}}
\affil{Korea Astronomy and Space Science Institute, Daejeon
305-348, Republic of Korea}

\author{Daisuke \textsc{Kuroda}}
\affil{National Institutes of Natural Sciences,
Okayama Astrophysical Observatory, Kamogata-cho, Okayama
719-0232, Japan}  

\author{Takafumi \textsc{Ootsubo}}
\affil{Astronomical Institute, Tohoku University, Aramaki,
Aoba-ku, Sendai 980-8578, Japan}


\author{Makoto \textsc{Sakamoto}, Shin-ya \textsc{Narusawa}, Jun \textsc{Takahashi}}
\affil{Nishi-Harima Astronomical Observatory, Center for Astronomy,
University of Hyogo, Sayo, Hyogo 679-5313, Japan}

\author{Hiroki \textsc{Akisawa}}
\affil{Himeji City Science Museum, Himeji, Hyogo 671-2222, Japan}

\and
\author{Jun-ichi \textsc{Watanabe}}
\affil{National Astronomical Observatory, 2-21-1 Osawa, Mitaka, Tokyo
181-8588, Japan}

\begin{abstract}
A Jupiter-family comet, 17P/Holmes, underwent outbursts in 1892 and
2007. In particular, the 2007 outburst is known as the greatest outburst
over the past century. However, little is known about the activity
before the outburst because it was unpredicted. In addition, the
time evolution of the nuclear physical status has not been
systematically studied. Here we study  the activity of 17P/Holmes
before and after the 2007 outburst  through optical and mid-infrared
observations. We found that the nucleus 
highly depleted its near-surface icy component before but became
activated after the 2007 outburst. Assuming a conventional
1-\micron-sized grain model, we derived a surface fractional active 
area of $0.58\%\pm 0.14\%$ before the outburst whereas it was enlarged by a
factor of $\sim$50 after the 2007 outburst. 
We also found that large ($\ge$1 mm) particles could be
dominant in the dust tail observed around aphelion. Based on the size of
the particles, the dust production rate was $\gtrsim$170 kg s$^{-1}$ at
the heliocentric distance $r_h=4.1$ AU, suggesting that the nucleus
still held active status around the aphelion passage. The nucleus color
was similar  to that of the dust particles and  average for a
Jupiter-family comet but different from most Kuiper Belt objects,
implying that the  color may be inherent to icy bodies in the solar
system.   On the basis of these results, we concluded that more than 76
 m surface materials were blown off by the 2007 outburst.
\end{abstract}

\keywords{interplanetary medium --- comets --- comets:
individual (17P/Holmes) --- solar system}

\section{INTRODUCTION}



17P/Holmes is a distinguished comet because of its spectacular outbursts
 in 1892 and 2007. There is no other  known comet  that has
exhibited such large-scale outbursts. However, its size and orbit are typical among Jupiter-family
comets: 17P/Holmes has a radius of 1.62--1.71 km
\citep{Snodgrass2006,Lamy2004} and orbital elements typical of the
Jupiter-family comets (e.g., a Tisserand parameter with respect to
Jupiter of 2.86) and most probably originated from the trans-Neptunian region
\citep{Levison1997}.  It passed through  perihelion at 2.05 AU on
 4.5 May 2007 and suddenly outbursted on 23 October 2007  at 2.44 AU from
the Sun in its outbound orbit.

Soon after the outburst, intensive research was conducted to
characterize its physical and chemical properties.
The optical magnitude data were compiled and the onset time of the
outburst was estimated to be 23.7
October \citep{Sekanina2008,Sekanina2009}. Later, \citet{Hsieh2010}
updated the onset time to $23.3\pm 0.3$ October based on  observations
obtained with SuperWASP-North.
\citet{Sekanina2008} deduced the duration of the active phase 
of the outburst to be $2.3\pm 0.3$ days. In contrast,
\citet{Altenhoff2009} indicated that the nucleus was activated by the
outburst and its activity continued for over 30 days. Small grains were
dominant near the nucleus after the outburst whereas large grains were
dominant gradually, probably because small grains were kicked out by
solar radiation pressure \citep{Zubko2011}. The mineralogical
properties of the dust grains were studied by mid-infrared
spectroscopy. It is found that the 
infrared spectrum of the diffuse emission can be explained by a
mixture of amorphous and crystalline silicate materials as observed
in most  comets \citep{Watanabe2009,Reach2010}. Total ejecta
mass was estimated to be 10$^{10}$--10$^{13}$ kg
\citep{Montalto2008,Altenhoff2009,Reach2010,Ishiguro2010,Boissier2012}. More
than a 2-m-thick surface layer could be blown off by the initial event
although this is  strongly dependent on the size distribution and
maximum size of the dust particles \citep{Ishiguro2010}. Cometary
fragments were also found by  optical wide-field
imaging. \citet{Stevenson2010} detected  16 fragments having a maximum
effective size of 10--100 m. The evidence of decameter-sized fragments
may suggest that $>$10 m surface layer could be excavated by the
outburst.   
\citet{Yang2009} detected two absorption features at 2 and
3 \micron~and suggested that the 17P/Holmes cloud contained a significant
fraction of pure water ice.

It is unlikely that the outburst was triggered by an impact of a small
object not only because the probability is incredibly low
\citep{Ishiguro2010} but also because of evidence that
multiple-outbursts occurred on the same comet in 1892
\citep{Barnard1896,Sekanina2008}. Several possible scenarios have been
presented to explain the cause of the outburst; these include   vaporization of
pockets of more volatile ices such as CO$_\mathrm{2}$ and CO
\citep{Schleicher2009,Kossacki2011}, the phase change of water from
amorphous to crystalline ice \citep{Sekanina2009}, thermal stress
in the nucleus, or the polymerization of hydrogen cyanide
\citep{Gronkowski2010}. 
So far, according to our present knowledge of the physical and chemical
structure of comets,  CO inclusions  are  likely to be the most probable
cause of Comet 17P/Holmes's outburst 
 \citep{Kossacki2010,Kossacki2011,Gronkowski2010}.

Despite the intensive observations soon after the explosion, however,
little is known about the change in activity between before and after
the outburst. It is important to know the nucleus's pre-outburst physical state  and how the nuclear surface changed through the
outburst. We found a unique pre-outburst data of 17P/Holmes
serendipitously obtained by AKARI, an infrared astronomy satellite, two
months before the 2007 outburst. In addition, we performed optical imaging over $\sim$3 years to monitor the long-term variation of
the comet's appearance and magnitude.
From our results, we confirmed that the comet was observed in a dormant
state before the 2007 outburst when it was near  perihelion whereas
it had  been active after the outburst even beyond 5 AU, exhibiting a
wide-spread dust tail. We describe the observations and  data reduction
in Section 2, and the observational results are presented in Section 3. 



\section{OBSERVATIONS AND DATA REDUCTION}

The data presented in this paper were taken with four ground-based
telescopes and one space infrared telescope: the Nishi-Harima
Astronomical Observatory (NHAO) NAYUTA 2.0-m telescope (hereafter
NHAO), the University of Hawaii 2.2-m telescope (UH2.2m), the Indian 
Institute of Astrophysics 2.0-m Himalayan Chandra telescope (HCT),
the Subaru 8.2-m telescope (Subaru), and the AKARI space infrared telescope. In
addition, we made use of data from the SMOKA archive, which is
operated by the Astronomy Data Center, National Astronomical Observatory
of Japan. All optical data
were taken more than one year after the outburst when the comet was
located at a heliocentric distance of 3.84  $\le$ $r_h$ $\le$ 5.17
AU. The infrared data from AKARI were obtained before the 2007 outburst
when the comet was at $r_h =2.23$ AU. A list of the observations is
given in Table 1. Details of the data acquisition and reductions 
are given in the following. 

\subsection{NHAO OBSERVATIONS IN   DECEMBER 2008 AND  JANUARY 2010} 
The Nishi-Harima Astronomical Observatory (134\arcdeg
20\arcmin08\arcsec E, 35\arcdeg 01\arcmin 31\arcsec N, 449 m) is a 
public astronomical observatory that conducts public relations
activities. Our observations were conducted as a part of the NHAO @site
program, which was contrived as a means of public outreach to introduce
visitors into cutting-edge astronomy through research experiences
\citep{Sakamoto2008}. We thus made observations with the @site program
participants. We employed MINT, a N$_2$-cooled optical CCD camera
mounted on the f/12 Cassegrain focal plane with a focal reducer and
$R_\mathrm{C}$-band filter. In this configuration, the pixel size 
on the sky was 0.276\arcsec, so that the field of view was
$9.4\arcmin\times9.4\arcmin$. The observations of 17P/Holmes were made using non-sidereal tracking on 23 and  26  December 2008  and 16
  January 2010 under photometric conditions. The average seeing size was 1.9\arcsec~on
23 December 2008, 2.0\arcsec~on 27  December 2008, and 1.6\arcsec~on
 16 January 2010, respectively. 

\subsection{HCT OBSERVATION ON  29--30 MARCH 2011}
The 2.0-m Ritchey-Chretien Himalayan Chandra Telescope  is located
in Hanle, India (78\arcdeg57\arcmin51\arcsec E, 32\arcdeg46\arcmin46\arcsec N,
4500 m), a high-altitude area in the Himalayan region. It is
operated by the Indian Institute of Astrophysics 
(IIA). We made observation on 29 March 2011 with the Himalaya Faint Object
Spectrograph (HFOSC) $2048\times 4096$ pixel CCD camera at the f/9
Cassegrain focus of the telescope. The telescope was controlled via a 
satellite from the CREST campus near Bangalore.
Half of the imaging area of HFOSC was sampled so that the effective
field of view was $10\arcmin\times 10\arcmin$ with a pixel scale of
0.296\arcsec. We made observation with an $R_\mathrm{C}$-band filter on
the first night while without filter on the second night. The observation
was performed under  photometric conditions. Point sources on the
images were spread out to 2.6\arcsec~probably because of inadequate
adjustment of the focus position or vibration of the telescope by
strong winds.

\subsection{SUBARU OBSERVATIONS ON  7 JANUARY  AND 5 JUNE  2011}

Subaru  is an 8.2-meter optical-infrared telescope at the
summit of Mauna Kea, Hawaii, operated by the National Astronomical
Observatory of Japan (NAOJ), National Institutes of Natural Sciences.
We carried out the observation of 17P/Holmes with the wide-field camera
Suprime-Cam \citep{Miyazaki2002} attached to the prime focus of the
8.2-m Subaru Telescope for two nights on 7 January 2011 and 5 June 2011. The camera has  a $34\arcmin\times 27\arcmin$ field of view with 
ten $2000\times 4000$ CCDs, whose pixel size is 0.20\arcsec. Since there are
gaps of a few arcseconds (over ten arcseconds between the CCD chips), we
dithered the telescope to make up for the gap areas. Five dithering modes
were applied for one complete set of images. We took images with an
$R_\mathrm{C}$-band filter on 7  January 2011 and with g$'$-, r$'$-, and i$'$-band
filters on 9 June 2011. These observations were carried out in a
non-sidereal tracking mode. The seeing size was 0.6\arcsec--0.8\arcsec.
The weather  was variable during the first run on 7 January 2011
and photometric conditions prevailed during the second run on 5 June 2011.

\subsection{UH2.2m OBSERVATION ON  4--5  FEBRUARY 2011}

The University of Hawaii 2.2-m observation was made for two nights on 4--5 February 
2011. We used a Tek2k and a Kron-Cousins
$R_\mathrm{C}$-band filter. The individual frames were taken in
non-sidereal tracking mode. The CCD was used in $1\times 1$ binned mode
on 4 February  and $2\times 2$ binned mode on 5 February.  The instrument
provides a $7.5\arcmin \times 7.5\arcmin$ field of view and a pixel
resolution of 0.22\arcsec~ (in $1\times 1$ binned mode) and 0.44\arcsec~
(in $2\times 2$ binned mode). At the time of the observation, the seeing
size was $\sim$1.0\arcsec. The weather conditions for these two nights
were photometric.

\subsection{AKARI ALL-SKY SURVEY}
AKARI (which was originally called  ASTRO-F), launched on 21  February 2006 UT, is a Japanese infrared
space telescope used to carry out an all-sky survey and pointed observations.
 It orbits at an altitude of $\sim$700 km
in a Sun-synchronous polar orbit along the boundary between night and day 
sides. The boresight vector of the telescope is pointed at the solar elongation
angle around 90\arcdeg~to suppress the incident thermal flux from 
the Earth and the Sun. AKARI consists of a bus module and a science module. 
The science module consists of a cryogenically cooled telescope of 68.5 cm 
aperture in diameter and two focal-plane instruments, the Far-Infrared Surveyor
(FIS) \citep{Kawada2007} and the Infrared Camera (IRC) \citep{Onaka2007}. 
Detailed descriptions on the design and operation of AKARI have appeared 
in \citet{Murakami2007}.
The all-sky survey, conducted between  8 May  2006 and  26 August  
2007,   is the major task of the AKARI project and the first half of  the mission period is dedicated
to it. 17P/Holmes  was
serendipitously detected with the longer channel of MIR in L18W
(13.9--25.3 $\mu$m) twice on 23 August 2007.

\subsection{DATA REDUCTION}

The observed optical raw data were reduced in the standard manner using
bias (zero exposure) frames recorded at intervals throughout the nights
plus skyflat data. The data were analyzed  using SDFRED2 for the  Subaru
data \citep{Ouchi2004} and IRAF for the other data. Flux calibration
was done using standard stars in the Landolt catalog
\citep{Landolt1992,Landolt2009} or field stars listed in the  USNO--B1.0
catalog \citep{Monet2003}. To convert pixel coordinates into celestial
coordinates, we employed the imcoords package in IRAF or  WCSTools.

To find the faint dust cloud structure, star-subtracted composite images
were produced in a method described in \citet{Ishiguro2007} and 
\citet{Ishiguro2008}. We first made images to align the stars to detect
faint stars and galaxies. We masked the identified objects using 
$\sim$3$\times $seeing-size circular masks. We also masked 
pixels identified as bad in the bias (hot pixels) and flat-fielding
images (pixels whose sensitivity was 5\% higher or lower than the
average). We combined the masked images with offsets to align 17P/Holmes,
excluding the masked pixels and shifting the background intensity to
zero. Since the comet moved relative to the stars, it was possible to
exclude nearly all masked pixels in the resultant composite image.

The AKARI infrared images was constructed in the same manner as
described in \citet{Ishihara2010}. The reduction pipeline process
includes a reset anomaly correction, a linearity and flat correction,
and  internal stray light removal. We applied the conversion factor
of 4.3 MJy/sr/ADU in L18W (D. Ishihara 2012, private communication).


\section{RESULTS AND DISCUSSION}

\begin{figure}
 \epsscale{0.7}
   \plotone{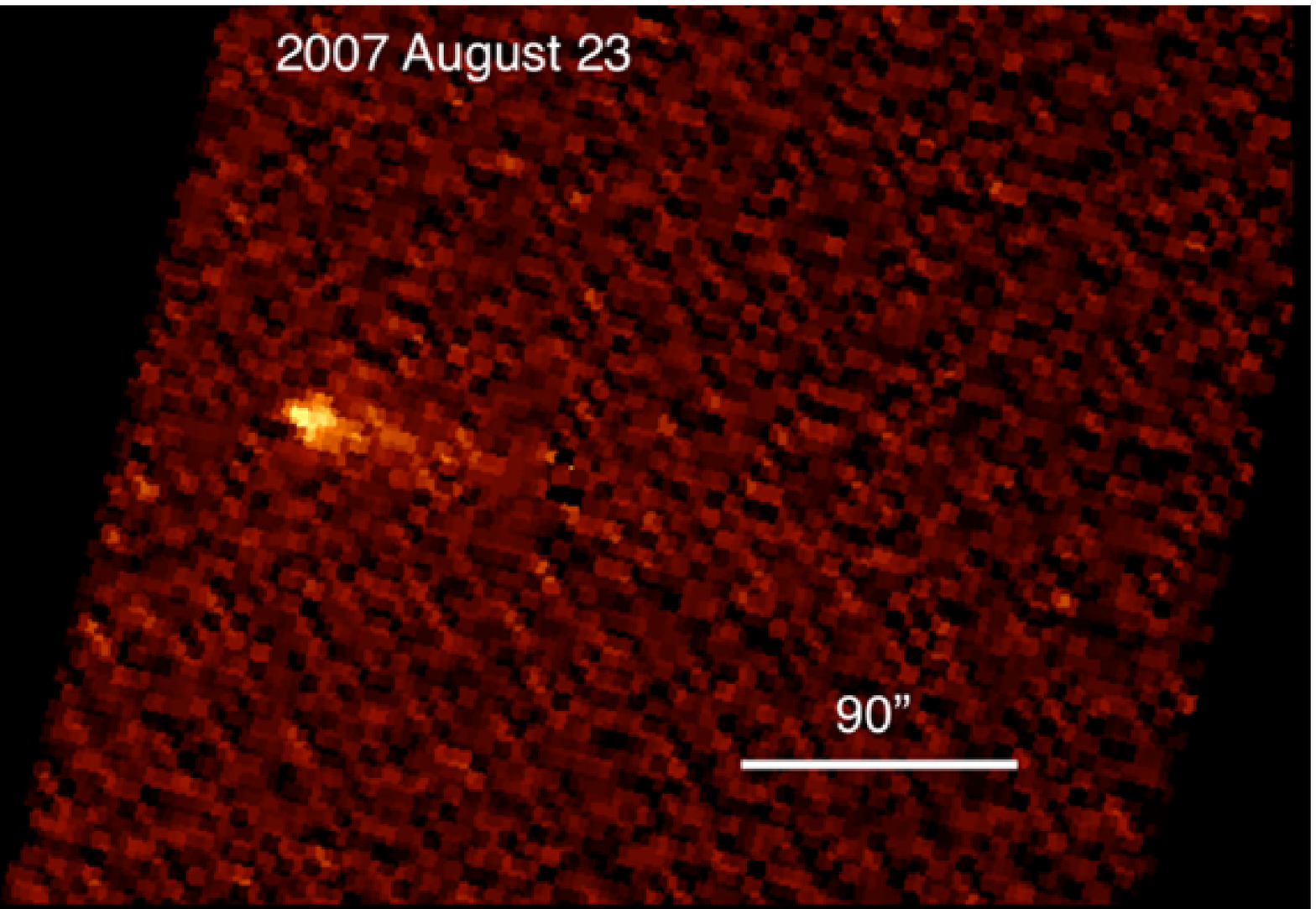}
  \caption{Mid-infrared pre-outburst image of 17P/Holmes taken on
 23 August 2007 with AKARI MIR.
 The image is the standard orientation in the sky, that is, Celestial North is
 up and East is to the left. The black areas correspond to the off-field
 region not covered with AKARI/IRC.}\label{fig:IRimage}     
\end{figure}

\begin{figure}
 \epsscale{1.0}
   \plotone{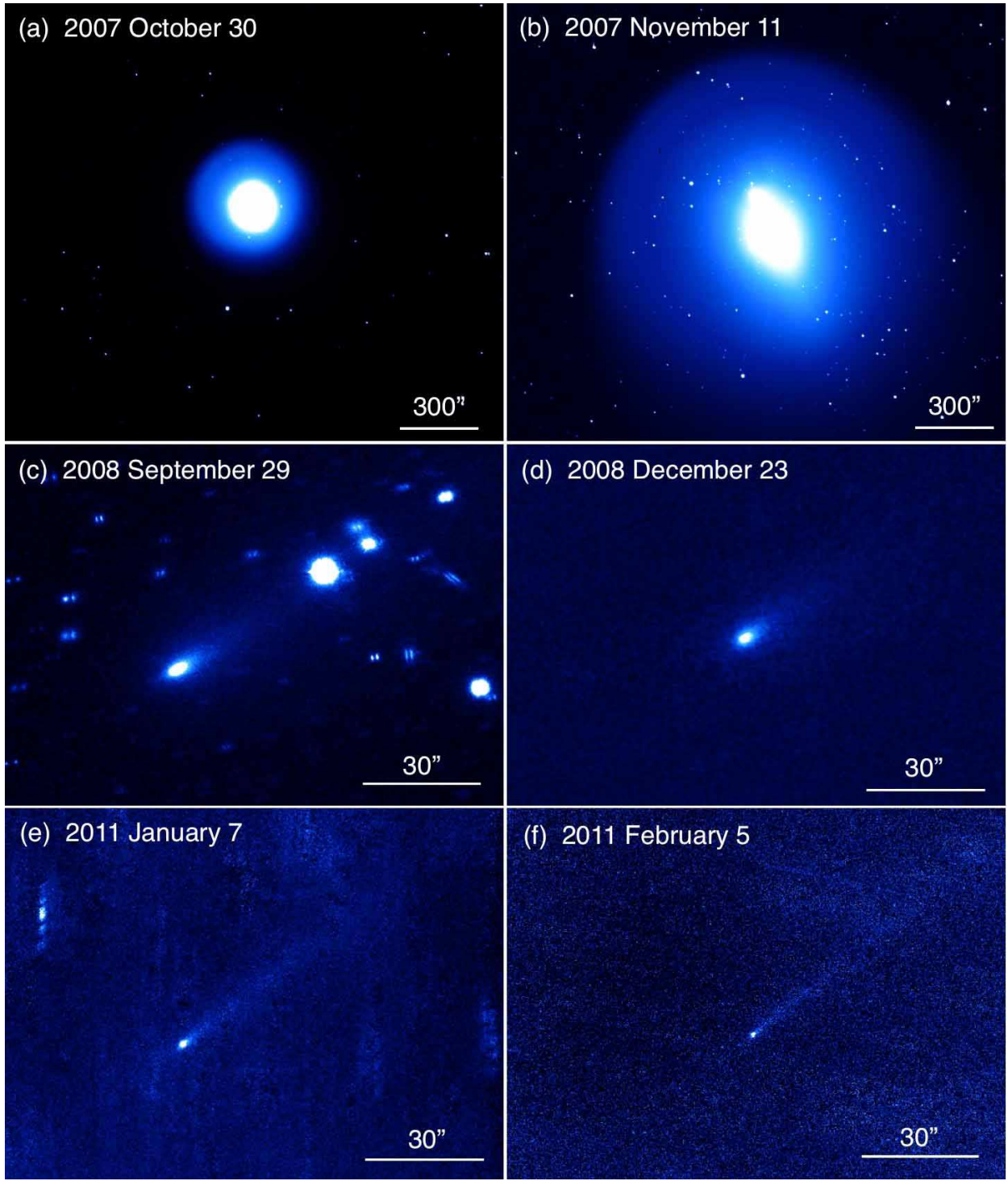}
  \caption{Composite optical images of 17P/Holmes taken after the
 2007 outburst. These images are the standard orientation in the sky,
 that is, Celestial North is  up and East is to the left. For reference,
 we show  two images soon  after the outburst from the SMOKA data
 archive (top two  images). 
 }\label{fig:OPTimage}     
\end{figure}

Figure \ref{fig:IRimage} shows the pre-outburst mid-infrared image of
17P/Holmes taken with AKARI on  23 August 2007. It consists of a near-nuclear dust coma and a faint tail extended toward the southwest.
Figure \ref{fig:OPTimage} shows the time-series post-outburst optical
images of 17P/Holmes between   30 October 2007  and   5 February 2011. For reference, images taken soon after the initial outburst are shown
in Figures \ref{fig:OPTimage}(a) and  (b); these were acquired with the Kiso
observatory 1.05-m Schmidt telescope and archived at SMOKA.
The dust cloud was initially observed as nearly spherical with
respect to the position of the nucleus (Figure \ref{fig:OPTimage}(a)),
and gradually it stretched toward the southwest (Figure \ref{fig:OPTimage}(b)). As it expanded, the inner coma becomes faint and pointlike
(Figures \ref{fig:OPTimage}(c)--(f)). 
Among these images, Figure \ref{fig:OPTimage}(e) is the most sensitive
among all of our data. Using the composite images with offset to align
stars, we estimated a detection limit of 26.7 mag in Figure
\ref{fig:OPTimage}(e).  Because there is no detectable fragment in the
image, we put the upper limit of the fragment radius at 400 m.
It is important to notice that all images in Figure \ref{fig:OPTimage}
show  the dust tail. Obviously, 17P/Holmes possessed a dust tail even
when it was located around  aphelion at 5.2 AU (Figures \ref{fig:OPTimage}(e)--(f)). However,  a pre-outburst observation on 3 May  and 3 June 3 2005  
revealed no comet-like coma at a heliocentric distance of
$r_h=4.66$ AU \citep{Snodgrass2006}. This fact may suggest that
an inactive dust layer was excavated by the outburst and fresh materials
of icy composition were exposed on the surface at the time
of our observations. In the following subsections, we provide a
quantitative analysis of the activities. 

\subsection{RADIAL PROFILES}

We examined  radial 
brightness profiles of the near-nuclear light source to confirm the
activity. We prepared  these profiles using AKARI data on
23 August 2007, NHAO data on 23 December 2008, UH2.2m data in 
February 2011 and Subaru data in  January 2011 and compared them with the
point spread function (PSF) determined by stars and asteroids
(Figure \ref{fig:radi}).  
Because the image was taken in non-sidereal tracking mode in the
optical, field stars were usually stretched out in the comet's
images. Infrared data were taken with short
exposure time ($<$1 s), so both the comet and background stellar objects remained
stationary in the observed data. 
Among the profiles in Figure \ref{fig:radi}, the Subaru data on 7 January 2011 is the best for the comparison because we set the individual exposure
time of 40 s, which is too short for the field stars to be
elongated in the non-sidereal tracking mode. In fact, the stars can be
extended no longer than 0.1\arcsec~ in Subaru data on 7 January 2011. For the AKARI data, we used the PSF of the average profiles of stars and
asteroids in the same scan pass. We determined the PSF of the other data
using images before or after the comet observations. They were usually
taken with short exposure times to confirm the position of 17P/Holmes
after the pointing of the telescopes. Therefore, there may be 
uncertainty in the time variation of the PSFs in NHAO data on 23 December
  2008 and UH2.2m data in  February 2011 (Figures \ref{fig:radi}(b) and (d)),
but probably the variations were $\lesssim$0.3\arcsec, which is typical
at these observational sites. 

In Figure \ref{fig:radi}, one can see that the surface
brightness profiles of 17P/Holmes were broader than those of point
sources at $\gtrsim$0.2\arcsec--1\arcsec~from the photometric center.
Hence we consider the epoch of the dust emission that is responsible
for the broadening of the radial brightness profiles.
The expansion dust speed has been studied in much of the literature. It depends
on how dust particles are coupled with 
the expanding gas molecules,  the expansion velocity of gas
molecules, the gas-to-dust mass ratio, and so on.
Surface orography and its inhomogeneities also play a significant role
in the terminal velocity of the grains \citep{Crifo1997}. 
Although there are many factors to determine the expansion dust
speed,  it can be approximated by a simple power-law function of the size and
the heliocentric distance, that is, $v_{ej} = $ $K / \sqrt{r_h a_d}$,
where $a_d$ denotes the radius of dust particles in micron. $K$ is a
constant, typically in the range of $100 < K < 1000$ m 
s$^{-1}$ based on theoretical studies \citep{Whipple1951,Ip1974}, past 
observations of normal comet activities
\citep{Lisse1998,Ishiguro2007,Sarugaku2007,Snodgrass2008,Ishiguro2008},
and cometary outbursts including the 17P/Holmes event in 2007
\citep{Sekanina2008,Montalto2008,Moreno2008,Hsieh2010,Sarugaku2010,
Reach2010,Stevenson2012}. From the equation for the expanding
speed, dust particles could escape from a region of 1\arcsec\  aperture
in  1.4--22 hours for $a_d=1$ \micron~and 6--94 days
for $a_d=1$ cm particles. Since our postburst data were taken
$>$342 days after the outburst, we can conjecture that the near-nuclear
dust particles were not the remnant of the particles ejected by the 2007
outburst. We discussed the staying time of the dust particles in
0.2\arcsec--1\arcsec~ aperture using a sophisticated dynamical model
(Appendix A), but both model results are consistent with each other in
that the near-nuclear dust particles were not the remnant of 2007
outburst. 


\subsection{PHOTOMETRY OF THE INNER DUST COMA}

We next deduced the dust coma magnitude as a
function of time and heliocentric distance. Photometry was performed
using the APPHOT package in IRAF, which  provides the magnitude
within synthetic circular apertures projected onto the sky. Since the
seeing disk sizes were different each night at each observatory, we set
the flexible aperture radius to 1.75\arcsec--6.50\arcsec, which
corresponds to 2.5  times the  full width at half maximum (FWHM). The sky
background was determined within a concentric  annulus having projected
inner and outer radii $2.5\times \rm FWHM$ and $3.0\times \rm FWHM$ of point
objects, respectively. The observed $R_\mathrm{C}$-band magnitudes,
$m_R$, are summarized in Table 2. The absolute magnitude,  the magnitude
at a hypothetical point at unit heliocentric distance and observer's 
distance, and at zero solar phase angle (Sun--object--observer's
angle),  is given by   
\begin{eqnarray}
m_R(1,1,0)=m_R - 5~\log(r_h \Delta) - 2.5\log \Phi(\alpha),
\label{eq:eq1}
\end{eqnarray}
\noindent where $\Delta$ is the observer's distance in AU and $\alpha$
is the solar phase angle in degree. The empirical
scattering phase function, $\Phi(\alpha)$, is given by the following
equation \citep{Lamy2004}:
\begin{eqnarray}
2.5\log \Phi(\alpha) = \beta \alpha ,
\label{eq:eq2}
\end{eqnarray}where $\beta$ characterizes the phase slope. $\beta=0.035$ mag
deg$^{-1}$ has been commonly assumed for cometary nuclei
\citep{Lamy2004,Snodgrass2006}. Alternatively, the phase function of
active comets is given based on observations of 1P/Halley given by
\citet{Schleicher1998}, \citet{Li2011}, and \citet{Stevenson2012},
\begin{eqnarray}
2.5\log \Phi(\alpha) = -0.045\alpha + 0.0004\alpha ^2 .
\label{eq:eq3}
\end{eqnarray}
We found that these different phase functions result in only less than a
1\%--2\% inconsistency at $\alpha=0$\arcdeg~ in the range of our
observations, i.e. $\alpha=3$\arcdeg--13\arcdeg. For this reason, we corrected
the phase angle dependence of the observed magnitudes using
Eq. (\ref{eq:eq2}) with $\beta=0.035$ mag deg$^{-1}$. 

Figure \ref{fig:mag} shows the absolute $R_\mathrm{C}$-band magnitude of
the dust coma as a function of the heliocentric distance. In the figure, we 
subtracted the nuclear magnitude, assuming 1.71-km spherical bodies and
a geometric albedo of 0.04 \citep{Lamy2004}. For comparison, we plot
the mean magnitude (the average of maximum and minimum magnitudes caused
by the rotating nucleus) obtained before the 2007 outburst
\citep{Snodgrass2006}, where the  authors could not find the detectable
coma. Figure \ref{fig:mag} clearly shows
that the photometric magnitudes after the outburst were significantly
brighter than the nuclear magnitude. This result is consistent with the
fact that the radial profiles of the near-nuclear light source were
broader than the stellar profiles (Section 3.1). In addition, there is a
trend that the magnitude decreased with increasing  heliocentric
distance. This trend can be attributed to the fact that the sublimation
rate of ice, which is responsible for the dust emission from the
nucleus, could decrease because of low solar flux. 
Accordingly, we can conclude that  17P/Holmes was active and had a faint
coma during the time of our observations. 

\subsection{DUST MASS-LOSS RATE }

Note that the absolute magnitude of active comets depends on the
aperture size for the photometry. A larger physical aperture encloses more
dust particles and accordingly the total cross section increases. 
\citet{Stevenson2012} used apertures of fixed physical radius at the
position of the comet to eliminate the effect. It is, however, difficult
to fix the physical aperture size for our data because the seeing sizes
differed at different sites from night to night. We 
adopted a method in \citet{Luu1992} to correct for the aperture size
effect. We converted the magnitude into the cross section and then mass-loss rate, by assuming that spherical dust grains with a certain radii and  mass
density. In addition, we utilized the infrared data from AKARI to derive
the dust mass-loss rate before the outburst in the manner described below.

We first calculated the cross section of coma dust particles, $C_c$,
and compared it with the area of the comet nucleus, $C_n$. We assumed
the 17P/Holmes nucleus to be spherical with a radius of 1.71 km 
\citep{Lamy2004}. We supposed that the scattering properties of 
the dust particles are the same as those of the nucleus because of the
small phase angles in our data set. We thus assumed a geometric albedo
of 0.04 and a scattering phase function given in Eq. (\ref{eq:eq2}) for
the dust particles. Second, we derived a 
parameter $\eta$ defined as  the ratio of the coma cross section $C_c$
to the nucleus cross section $C_n$. At optical wavelengths, $\eta$  is proportional to the ratio of the flux density scattered
by the coma, $I_c$, to the flux density scattered by the nucleus cross
section, $I_n$, which enables us to characterize the contribution of dust
particles in the coma \citep{Luu1992,Hsieh2005,Kasuga2008}, that is,
\begin{eqnarray}
\eta \equiv \frac{C_c}{C_n} \simeq \frac{I_c}{I_n}~~~(\mathrm{optical}).
\label{eq:eq11a}
\end{eqnarray}

For the mid-infrared data, it is improper to use Eq. (\ref{eq:eq11a}) for
the derivation of $\eta$ because the thermal properties of dust particles
are largely different from those of cometary nuclei. We modified
Eq. (\ref{eq:eq11a}) into
\begin{eqnarray}
\eta \equiv \frac{C_c}{C_n} =
 \frac{I_c}{I_n}\left(\frac{i_c}{i_n}\right)^{-1}~~~(\mathrm{infrared}), 
\label{eq:eq11b}
\end{eqnarray}

\noindent where $i_n$ is the flux from a big spherical body like
a comet nuclei of unit cross-sectional area. We calculated
the flux by using the standard thermal model (STM) \citep{Lebofsky1989}. In the
model, it is assumed that the nucleus is a nonrotating spherical
body. We thus considered that each element of the surface is in
instantaneous equilibrium with solar influx. In situ observation with the
Deep Impact spacecraft revealed that the STM was a good approximation to
characterize the thermal balance of a comet nucleus whose surface
consists of dry materials without icy components
\citep{AHearn2005}. Standard thermal parameters are assumed, i.e., 
emissivity $\epsilon_E=0.90$--0.95,  beaming parameter
$\eta_E=0.756$--0.850,  phase integral $q_E=0.28$--0.75, and thermal
phase coefficient $\beta_E=0.01$ mag deg$^{-1}$. In the range, we
obtained $i_n=(5.3\text{--}6.0)\times 10^{-10}$ Jy/m$^2$. In contrast,
the flux from the dust coma, $i_c$ that has an equivalent total cross-sectional  area was derived in the manner described in
\citet{Ishiguro2010}.  We calculated the equilibrium temperature 
of a 1-\micron-sized particle $r_h$=2.23 AU using the optical constants
of astronomical silicate (217 K), magnetite (229 K), and graphite (235
K) and derived a thermal flux at 18-\micron~AKARI wavelength of
$i_c=2.0\times 10^{-9}$ Jy/m$^2$ for astronomical silicate,
$i_c=2.4\times 10^{-9}$ Jy/m$^2$ for magnetite, and
$i_c=2.6\times 10^{-9}$ Jy/m$^2$ for graphite. We obtained the observed
flux $I_c+I_n=0.39\pm 0.04$ Jy. Using the STM, we estimated the flux from
the nucleus to be $I_n=(4.8$--5.5)$\times$10$^{-2}$  Jy. Substituting these values in
Eq. (\ref{eq:eq11b}), we obtained $\eta=17\pm 4$ for the pre-outburst
data. The derived $\eta$ values are summarized in Table 2.


The dust mass-loss rate can be derived from $\eta$ in the manner in
\citet{Luu1992} as 
\begin{eqnarray}
\dot{M}_d = \frac{1.1\times10^{-3}\pi ~\rho_d ~\bar{a}
 ~\eta~r^2_{obj}}{\phi ~r_h^{1/2} ~\Delta},
\label{eq:eq12}
\end{eqnarray}
\noindent where $\rho_d$ is the mass density of the dust
particles, $\bar{a}$ is the grain radius in meters, $r_{obj}$ is the
radius of the 17P/Holmes nucleus, and $\phi$ is the reference photometry
aperture radius in arcsec. We assumed that 17P/Holmes emitted small dust
grains, that is, $\bar{a}=1.0 \times 10^{-6}$ m.  We supposed  the
mass density of dust particles to be  $\rho_d=1000$ kg m$^{-3}$. The derived
dust mass-loss rates are listed in Table 2. Note here that the dust
mass-loss rate is  a crude estimate. In fact, there is a big uncertainty
in the dust mass-loss rate because the mass-loss rate is proportional to
the grain size.

Figure \ref{fig:Qdust} shows the dust mass-loss rate as a function of
 heliocentric distance $r_h$. In addition to our data, we compared
the dust mass-loss rate to data from  previous research. 
As we previously mentioned, \citet{Snodgrass2006} could not detect any coma
at 4.66 AU before the outburst and put an upper limit on an unresolved coma
of 24.6 mag. We converted the magnitude into the mass-loss rate in the
figure. \citet{Miles2010} monitored the near-nuclear magnitude over five months using 2.0-m telescopes, the
Faulkes Telescope North and the Liverpool Telescope, with an SDSS-r$'$
filter, and derived the magnitude. Since the aperture size for
photometry and the bandpass filter of their observation were different
from ours, we scaled their data to match our data at the NHAO run in December 2008. Owing to  frequent observations as well as to good
photometric stability, \citet{Miles2010} succeeded 
in  detecting a minor burst possibly occurring on  $4.7\pm 0.5 $ January 2009 and
 attaining a peak magnitude enhancement of $0.85\pm 0.1$ mag.
Moreover, we refer to the dust mass-loss rate derived in
\citet{Stevenson2012}. The observation covered when the comet was at
2.49--2.50 AU. They derived the mass-loss rate in a manner similar to ours.
They thus obtained the mass-loss rate by assuming spherical dust grains with
radii of 1 \micron~ and a bulk density of 1000 kg m$^{-3}$.

In Figure \ref{fig:Qdust}, it is clear that the mass-loss rate
significantly increased after the 2007 outburst. In addition, it
had decreased with increasing  heliocentric distance most likely
because of weaker solar influx. There are two minor peaks on 12 
November 2007 and 5 January  2009, indicating minibursts occurred at those
epochs \citep{Stevenson2012,Miles2010}. 

\subsection{FRACTIONAL ACTIVE AREA OF THE NUCLEUS}
We now consider a model to predict the dust mass-loss rate based
on a thermal balance on the surface.
We assume that this element of the surface is in instantaneous
equilibrium with solar radiation and the latent heat of sublimation of
ice. We thus consider the energy balance on the surface of the icy
body given by
\begin{eqnarray}
\frac{S_0}{r_h^2}(1-A_p) \cos ~z=\epsilon_E \sigma T^4 + L_w(T)\frac{dZ}{dt}~~
 \label{eq:eq13}
\end{eqnarray}
\noindent \citep{desvoivres1999,desvoivres2000},
where $S_0$ is the solar flux at 1 AU, $z$ is the zenith distance of the
Sun, $\epsilon_E$ is the emissivity, and $\sigma$ is the Stefan-Boltzmann
constant. $A_p$ is the geometric albedo and $T$ is the surface temperature.
The latent heat of sublimation of water, $L_{w}$ is given by
\begin{eqnarray}
L_w = 2.886 \times 10^6 - 1116~T~~\mathrm{J~kg^{-1}}.
\label{eq:eq14}
\end{eqnarray}
The sublimation rate of the water ice is given by
\begin{eqnarray}
\frac{dZ}{dt}~=~\frac{1}{1+1/\kappa}~\gamma(T)~P_w(T)~\sqrt{\frac{m_w}{2\pi
 k T}} ~~\mathrm{kg~s^{-1}},
 \label{eq:eq15}
\end{eqnarray}
\noindent
where $\kappa$ is the water ice-to-dust mass ratio, defined as $\kappa$
= $\rho_w/\rho_d$ (where $\rho_w$ and $\rho_d$ are the masses of water ice and
dust particles per unit volume, respectively), $m_w$ is the molecular
mass of  water, and $k$ is the Boltzmann constant. $\gamma$ denotes the
sticking coefficient \citep{haynes1992,enzian1997} given by
\begin{eqnarray}
\gamma(T)~=~-2.1 \times 10^{-3} ~T + 1.042 ~~~(T>20~\mathrm{K}). 
 \label{eq:eq16}
\end{eqnarray}

In Eq. (\ref{eq:eq15}), the saturated vapor pressure of water, $P_w(T),$
is given by the Clausius-Clapeyron equation. The mass-loss rate of the
dust particles, $\dot{M}_d $, is therefore given by an integral over the
sunlit hemisphere of the spherical body:
\begin{eqnarray}
\dot{M}_d~=~\frac{2 \pi r_{obj}^2 f }{\kappa}~
 \int_0^{\pi/2} \left(\frac{dZ}{dt}\right) ~\sin z \,dz ~~\mathrm{kg},
 \label{eq:eq17}
\end{eqnarray}

\noindent
where $f$ is the fractional active area. We used $\epsilon_E=0.9$.

In this model, there is a large uncertainty in $\kappa$. It is conventionally
assumed to be unity in the literature. Once we fix $\kappa$, it is possible
to determine $f$ by a comparison with the observed mass-loss rate. We
considered five different cases in the range of 0.1$\le$ $\kappa$
$\le$10 and fit the model results to the observed data at 4.1 AU,
adjusting $f$. Figure \ref{fig:compQdust} shows the heliocentric distance
dependence of the dust mass-loss rate. In the figure, we compared the
observed dust mass-loss rate  with those calculated from the models.
Our models reproduce the trend that the dust mass-loss rate decreases with
increasing  heliocentric distance. However, there are big differences
at $<$3 AU and at the time of the miniburst on 5 January 2009. We infer that
there could be  remnants of large dust particles in the physical aperture
 \citep{Stevenson2012} or that the nucleus was still in an extraordinarily
excited state in the aftermath of the 2007 outburst.

Infrared space observations of periodic comets suggested that the
mass-loss rate of dust trail particles is comparable to that inferred
from OH production rates or 
larger than that inferred from visible-light scattering in comae
\citep{Sykes1992,Reach2007,Lisse2006}. A theoretical model to simulate
the recurrent outburst nature of  the 17P/Holmes outburst shows  favorable
results when $\kappa=0.4$--0.6 \citep{Hillman2012}. MODEL 2
($\kappa=0.32$) and MODEL 3 ($\kappa$=1.00) are therefore reasonable
models among these five. 
When we adopt these models, we can derive the fractional active area, as
shown in Figure \ref{fig:frac}. The fractional active area was 0.20--0.38
over the time of our observations but increased  to 0.64 when the
miniburst occurred. The $f$ values are larger than the
average for short-period comets ($<$0.2 on average; see,
e.g., \citet{Tancredi2006}). Note again that the dust mass-loss rate
derived from the  observation strongly depends on the particle size. If
we assume 10-\micron-sized particles, the fractional active area is
saturated to be unity. Therefore, we conclude that a significant
fraction of the surface of 17P/Holmes was still active. 

\subsection{DUST TAIL MORPHOLOGY}

So far, we derived the dust mass-loss rate under the assumption of small
particles (i.e., $a_d=1$ \micron). The model has been widely used
in previous research \citep[see, e.g.,][]{Luu1992}. It also permits a
direct comparison with the previous study of the 17P/Holmes dust mass-loss rate
given in  \citet{Stevenson2012}. However, the existence of large particles
as well as micrometer-sized particles is widely confirmed based on
telescopic observations 
\citep{Watanabe1990,Ishiguro2002,Sykes1992,Fulle2004,Reach2007},
remote-sensing observations with spacecraft onboard cameras
\citep{Sekanina2004}, and in situ measurements of cometary dust particles
\citep{Mcdonnell1986,Tuzzolino03,AHearn2011}. The detection of a cometary
dust trail associated with 17P/Holmes with Spitzer is definitive
evidence that 17P/Holmes had ejected large dust particles
\citep{Reach2010}. Furthermore, \citet{Moreno2008} suggested that $>$600
\micron~ particles could be ejected by the 2007
outburst. Millimeter-wavelength continuum observations also suggested 
the existence of submillimeter particles \citep{Altenhoff2009}. We hence
examined the dust particle size using our observed composite images.

Orbits of dust grains are, in principle, determined by the size and the
ejection speed. The size of the particle can be parametrized by
$\beta_{rp}$, the ratio of the solar radiation pressure to the
gravitational attraction. Assuming a spherical particle, we can define $\beta_{rp}$  as
\begin{eqnarray}
\beta_{rp} = \frac{K~ Q_{pr}}{\rho_d~ a_d},
\end{eqnarray}
\noindent where $a_d$ and $\rho_d$ are the particle radius in
meters and the mass density in kg m$^{-3}$. $K$ = 5.7 $\times$ 10$^{-5}$ kg
m$^{-2}$ is a constant and $Q_{pr}$ is the radiation pressure coefficient
averaged over the solar spectrum \citep{Burns1979}. Supposing that
particles are compact in shape and large compared to the optical
wavelength, we considered $Q_{pr}$ as unity.

As an initial guess, we drew the syndyne and synchrone curves. Syndynes
are curves representing a constant value of $\beta_{rp}$ when the ejection
velocity is assumed to be zero. Synchrones are curves representing the
positions of particles of different sizes (i.e., different
$\beta_{rp}$) ejected at the same time with zero velocity.
Figure \ref{fig:sync-synd} left shows a comparison between the observed
contour maps and the syndyne curves of $a_d=1$ \micron, 10 \micron,
100\micron, 1 mm, and 1 cm. We selected two composite images on 23
December 2008 and 7 January 2011 because synchrones and syndynes are
well separated in these images. From the syndyne curves, the loci of
larger particles, $a_d=0.1$--1 cm, match the center of the dust
tail. This suggests that big particles are dominant in the
cross-sectional area in the dust tail. From the synchrone curves,
particles ejected more than one year before the observation match  the
position of the dust tail (Figure \ref{fig:sync-synd} right). It is
likely that the dust particles ejected soon after the outburst were
responsible for the dust tail. Synchrone--syndyne analysis, however,
tends to lead to a misleading value of dust sizes
\citep{Ishiguro2007,Fulle2004}. A three-dimensional analysis, which
allows nonzero ejection velocities to be considered, is appropriate to 
estimate the particle sizes and mass-loss rate. Here we applied a
three-dimensional analysis to match the observed images, following the model
in \citet{Ishiguro2007}, \citet{Sarugaku2007}, \citet{Ishiguro2008}, and
\citet{Hanayama2012}. 

We assumed that the dust particles were ejected  symmetrically with
respect to the Sun--comet axis in a cone-shape jet  with half-opening
angle $w$, implying that the active regions are distributed ubiquitously over
the surface of the nucleus and therefore that the dust emission occurred
homogeneously around the subsolar point. The model also suggests that
the dust particles were ejected independently of  the rotation of the
nucleus. Given that the ejection speed was a power-law function of
heliocentric distance, we adopted an empirical function for the ejection
terminal velocity of dust particles: 
\begin{equation}
V_{ej} = V_0 \left(\frac{\beta_{rp}}{\beta_{rp,0}}\right)^{u_1}\left(\frac{r_h}{r_0}\right)^{-u_2}
v
,\end{equation}

\noindent
where $V_0$ is the reference ejection velocity of the $\beta_{rp}=\beta_{rp,0}$
particles at $r_h=r_0$. We set $\beta_{rp,0}=1$ and $r_0=1$ AU,
respectively. $u_1$ and $u_2$ are the power indices of  $\beta_{rp}$ and
the heliocentric  distance $r_h$ dependence of the ejection velocity.
A random variable $v$ follows the Gaussian probability density function,
$P(v)$, which is given by the following formula:
\begin{equation}
P(v) = \frac{1}{\sqrt{2\pi}\sigma_v}\exp \left(- \frac{(v-1)^2}{2\sigma_v^2}\right)    .
\end{equation}
where $\sigma_v$ is standard deviation of $v$. We set $\sigma_v$=0.25. 
From this method, most $v$ (68\%) has a value ranging from 1-$\sigma_v$
to 1+$\sigma_v$.

A power-law size distribution with index $q$ was used, i.e.,  
\begin{equation}
N(a;t)~da~dt~=N_0 \left(\frac{r_h}{r_0}\right)^{-k}
 \left(\frac{a_d}{a_{0}}\right)^{-q}~da\,dt , 
\end{equation}
\noindent
in the size range of $a_{min}$ $\le$ $a_d$ $\le$ $a_{max}$. 

We derived the above parameters to fit the morphology of the dust cloud.
We focused on the positions of the extended dust tail and the flux
ratio between coma and extended structures. 
The input and best parameters sets are summarized in Table 3. We show an
example simulation image to compare with the observational data on 23
December 2008. From the fitting we derived a mass-loss rate at 4.1 AU of
$>$170 kg s$^{-1}$. Although there are large uncertainties in these
best-fit values, it is clear that the resultant mass-loss rate is
significantly larger than the rate we derived based on the small-grain
model in Section 3.3.

\subsection{TOTAL MASS AND DEPTH EXCAVATED BY THE OUTBURST}
We hereby consider the total mass of the ejecta and the depth
excavated by the 2007 outburst. There is a big uncertainty in the total
mass in the range of 10$^{10}$--10$^{13}$ kg
\citep{Montalto2008,Altenhoff2009,Reach2010,Ishiguro2010,Boissier2012}.
In addition, little is known about the excavated depth by the initial
outburst. As discussed in \citet{Ishiguro2010}, the power index of the
size distribution and the maximum size of the particles are critical
factors to derive the total mass. The fraction of 
the active area is a crucial factor in determining the depth. In the
previous subsection, we confirmed that 1mm--1cm particles were ejected
from the nucleus. The power index of the size distribution was derived
to be 3.4--3.6, which is consistent with previous studies
\citep{Zubko2011,Boissier2012}. Moreover, we deduced the fractional
active area of 0.20--0.38. It is reasonable to think 20--38\% of the
surface materials were blown out by the initial outburst if the surface
condition remained constant after the outburst. Using these parameters
and assuming the mass density of the particles and the nucleus are
$\rho_d$=1 g cm$^{-3}$ and $\rho_n$=0.5 g cm$^{-3}$, respectively, we
could update the total mass and newly derive the depth excavated by the
2007 outburst in the same manner as \citet{Ishiguro2010}. We found that
the total mass of the ejecta was in the range of 5.3$\times$10$^{11}$--
6.1$\times$10$^{12}$ kg. The derived mass is consistent with
\citet{Reach2010} and \citet{Boissier2012}. In addition, it was found
that 76m--1600m of surface materials were excavated by the initial
outburst. Since the upper limit of the depth, 1600 m, is equivalent to
the  nuclear radius, our upper limit may not be a realistic
estimate. Therefore, we can safely conclude that more than 76 m surface
materials were blown off by the 2007 outburst. 

\subsection{DUST COLOR}
We measured the color of the dust within an aperture radius of
1.0\arcsec~plus the nucleus using Subaru data on 6 June 2011. The
observed magnitudes are $23.63\pm 0.10$ (g$'$ band), $23.15\pm 0.09$
(r$'$ band), and $22.64 \pm 0.08$ (z$'$ band), respectively. The photometric
magnitudes were converted into VRI magnitudes using  transformation
equations from Smith et al. (2002). We derived the color indices
$V-R=0.39\pm0.08$ and $R-I=0.54\pm0.07$. These color indies are consistent
with those of the nucleus before the outburst, that is,
$V-R=0.41\pm0.07$ and $R-I=0.44\pm0.08$ \citep{Snodgrass2006}.
They are also similar to the average of Jupiter-family comets but different
from most Kuiper Belt objects \citep{Jewitt2002,Meech2004}. Because the old
surface was excavated and fresh surface was exposed in our data, the
color may be inherent to fresh comets, that is, icy bodies in the solar
system.

\section{SUMMARY}

We have outlined the observational evidence of 17P/Holmes
activity before and after the 2007 outburst and found the followings:

\begin{itemize}
\item The nucleus highly depleted its near-surface icy component
      before the 2007 outburst.
\item It had been active even near the aphelion passage in 2010.
\item The surface fractional active area was $0.58\%\pm0.14\%$ before the
      outburst whereas it enlarged by a factor of $\sim$50 after the 2007
      outburst under the assumption of a small-dust-grain model.
\item The nucleus color was similar to that of the dust particles and to the
      pre-outburst color of the nucleus.
\item More than 76 m surface materials were blown off by the 2007 outburst.
\end{itemize}

We expect that long-term monitoring observations during the 2014
perihelion passage and even later will give important information about
how the active nucleus depletes its near-surface ice.

\vspace{1cm}
{\bf Acknowledgments}\\
We express our gratitude to the participants of the @site program for
helping with our observations. We shared a joyful occasion with the
public visitors at NHAO.  Use of the UH2.2m telescope was supported by
NAOJ, and HCT observation was assisted by the observatory staff at the
Indian Institute of Astrophysics. Subaru data were obtained with the
support of  the National Astronomical Observatory. We particularly thank 
Drs. Fumiaki Nakata and Miki Ishii for their  observational
support and technical advice.  The infrared data is acquired with AKARI,
a JAXA project with the participation of ESA. A portion of the data was
collected with the Subaru Telescope and obtained from  SMOKA, which is
operated by the Astronomy Data Center, National Astronomical Observatory
of Japan. We thank the referee for his/her careful reading and valuable
comments. SH is supported by the Space Plasma Laboratory, ISAS/JAXA.
This work at Seoul National University was supported by a National
Research Foundation of Korea (NRF) grant funded by the Korean government
(MEST) (No. 2012R1A4A1028713).

\appendix
\section{CONTRIBUTION OF DUST PARTICLES EJECTED THROUGH 2007 OUTBURST}
\begin{figure}
 \epsscale{1.0}
   \plotone{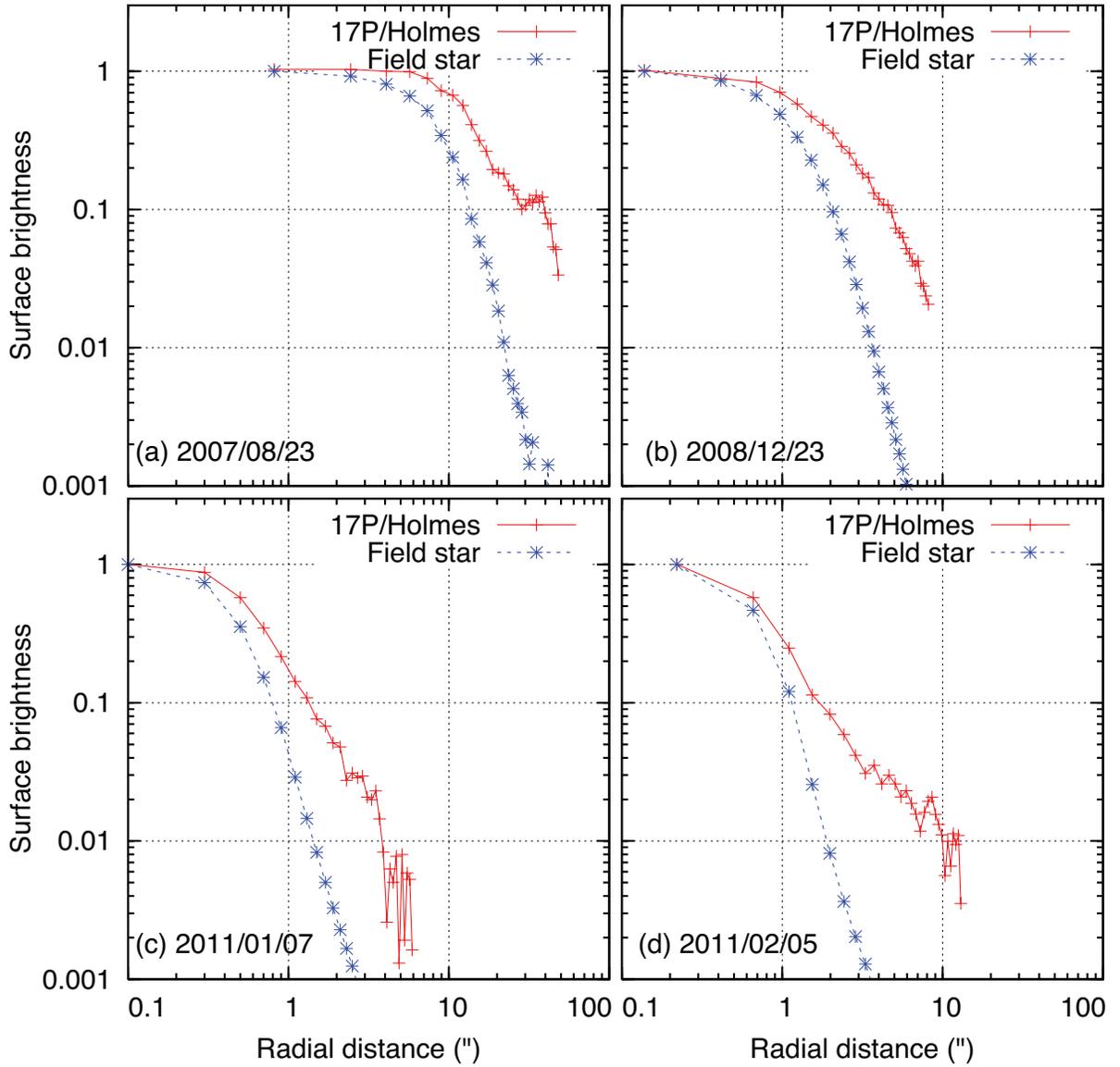}
  \caption{Surface brightness profiles of 17P/Holmes, normalized to
 the brightness at 0.1\arcsec~from the nucleus position, for composite
 images from observations made on (a) 23 August 2007, (b) 23  December 2008, (c) 7 January 2011, and (d) 5  February 2011. Note that (a) is
 obtained  at mid-infrared wavelength with AKARI whereas the others are at the
  optical wavelength with ground-based telescopes.}\label{fig:radi} 
\end{figure}

\begin{figure}
 \epsscale{1.0}
   \plotone{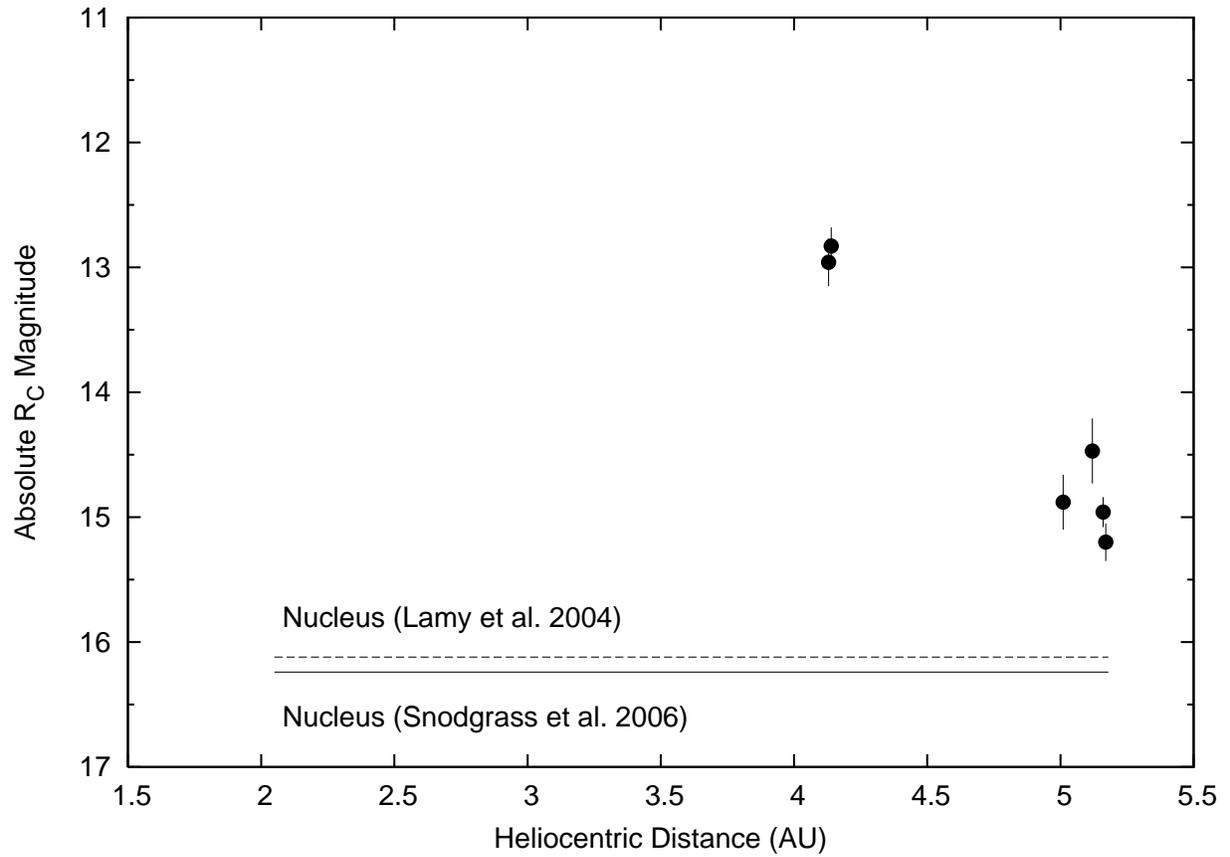}
  \caption{Heliocentric distance dependence of the absolute $R_\mathrm{C}$
 magnitude. For comparison, we show the predicted magnitude of
 a 1.71-km-sized nucleus with a geometric albedo of 0.04
 \citep{Lamy2004} and the pre-outburst magnitude in
 \citet{Snodgrass2006}.}\label{fig:mag} 
\end{figure}

\begin{figure}
 \epsscale{1.0}
   \plotone{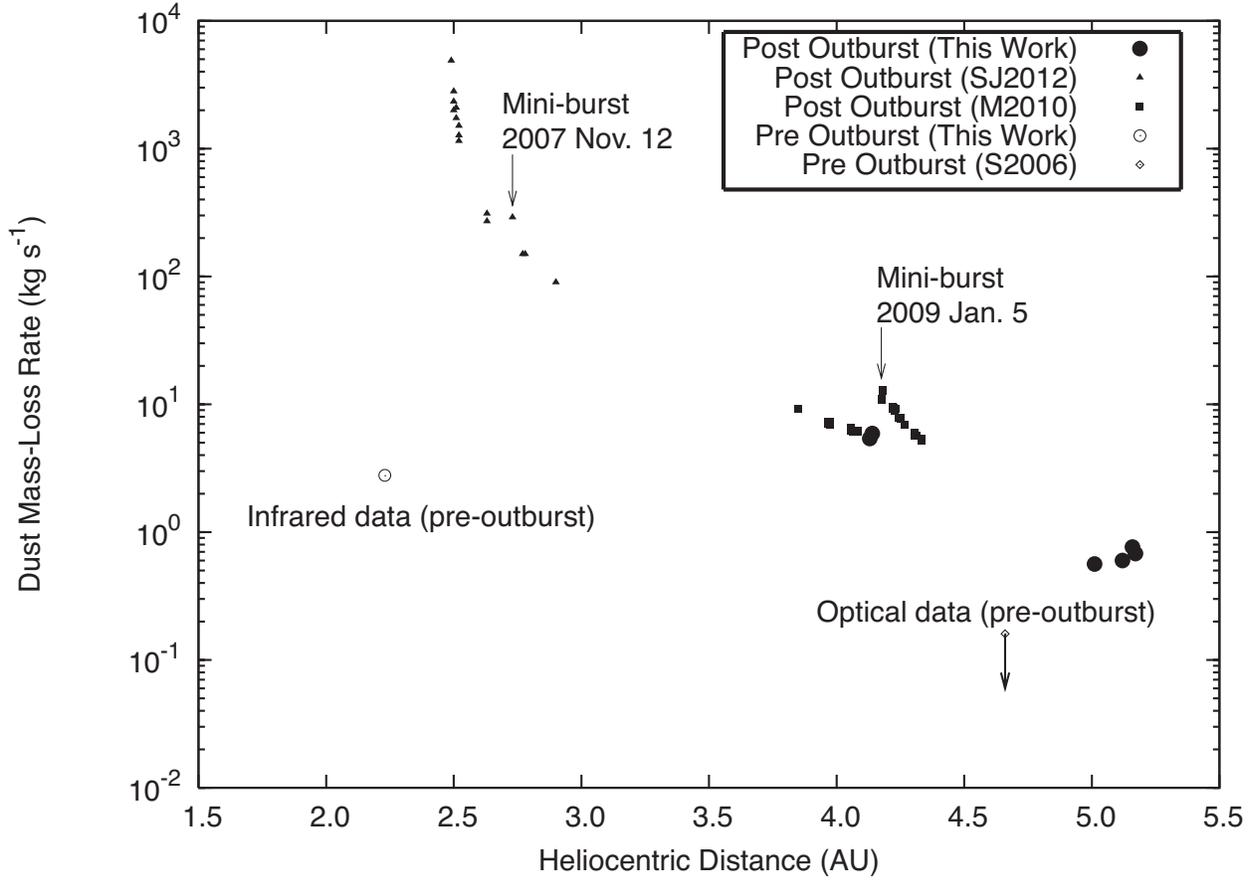}
  \caption{Heliocentric distance dependence of the dust mass-loss rate.
 For comparison, we show the dust mass-loss rate derived in
 \citet{Stevenson2012} (SJ2012 in the figure). The magnitudes in
\citet{Miles2010} were  calibrated to match our data and  used to
derive the dust mass-loss rate at $r_h=3.85$--4.33 AU (M2010). The upper
limit of the dust coma magnitude in  \citet{Snodgrass2006} was used to
derive the upper limit for the dust mass-loss rate before the outburst
 at $r_h=4.66$ AU (S2006). Two thin arrows 
 indicate the possible minibursts occurring on  12 November  2007
 \citep{Stevenson2012} and 5 January 2009 \citep{Miles2010}, respectively.
}\label{fig:Qdust} 
\end{figure}

\begin{figure}
 \epsscale{1.0}
   \plotone{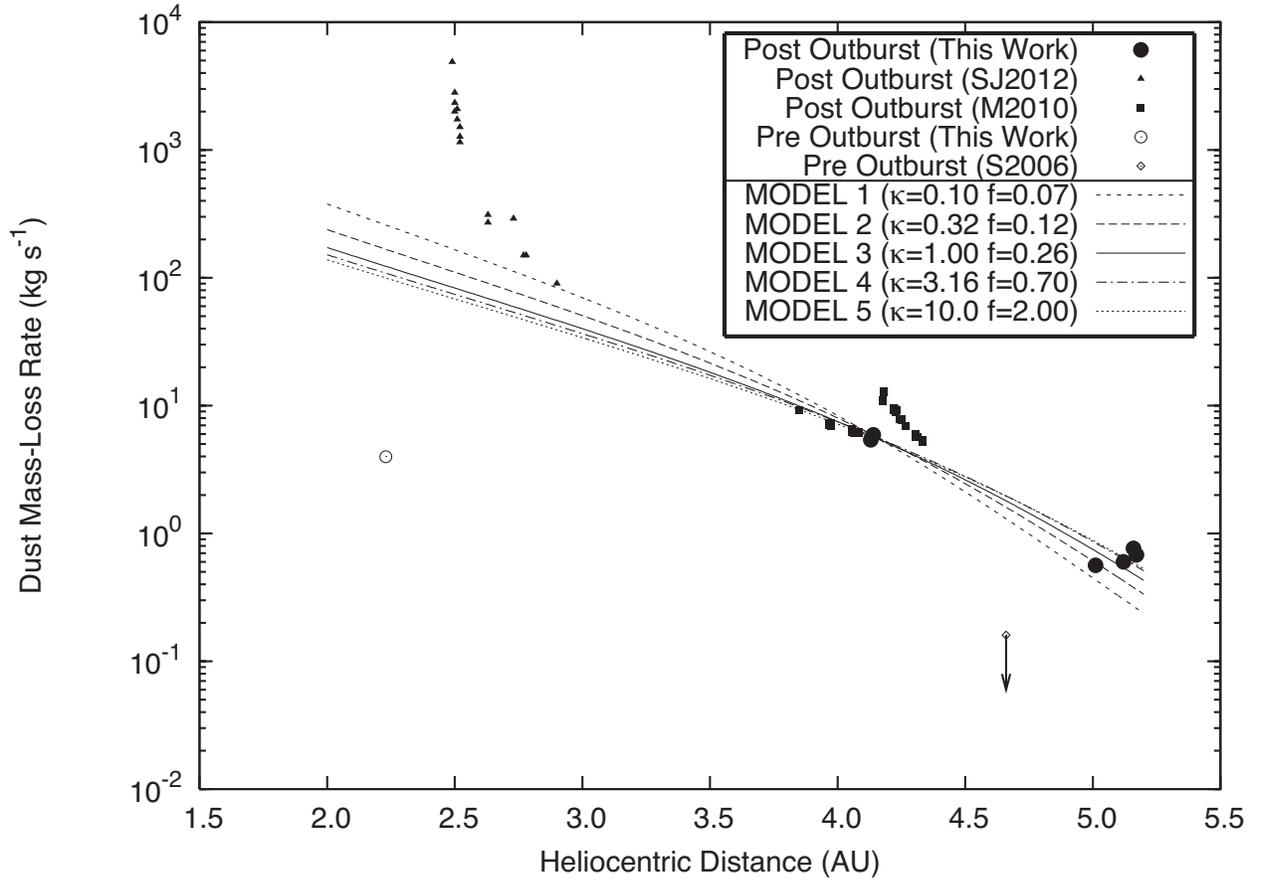}
  \caption{Comparison of the dust mass-loss rate between observations
 and models. }\label{fig:compQdust} 
\end{figure}

\begin{figure}
 \epsscale{1.0}
   \plotone{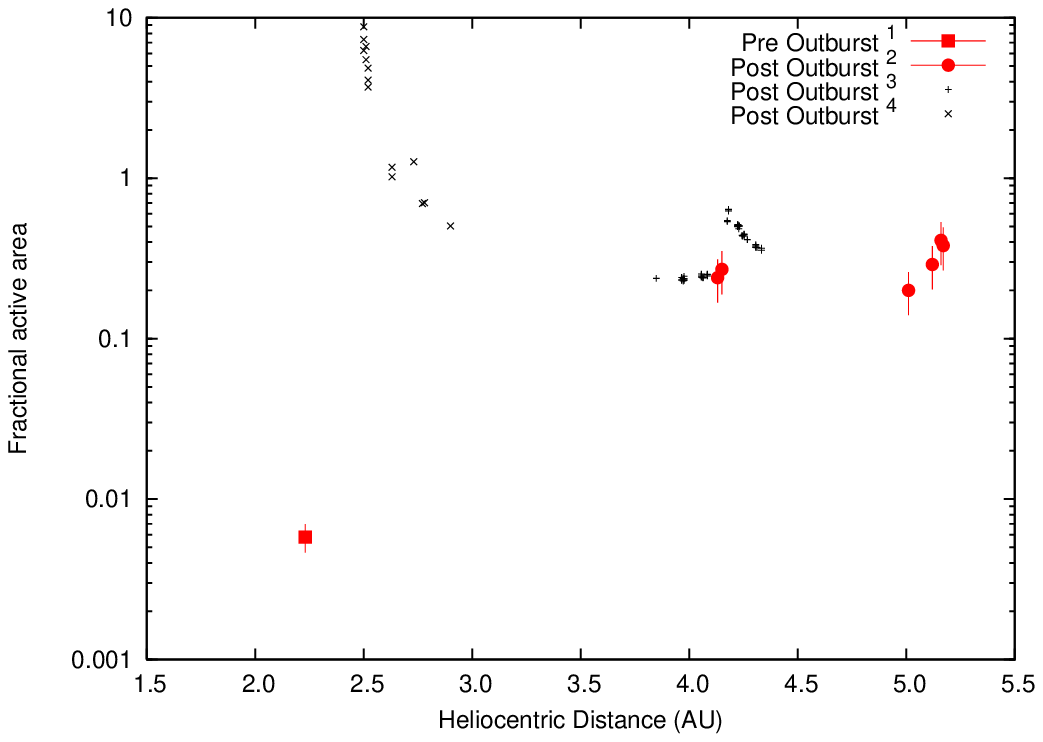}
  \caption{Fraction of active area of the 17P/Holmes nucleus as a function
 of the heliocentric distance. In this figure, we assumed the water
 ice-to-dust mass ratio $\kappa=1$.}\label{fig:frac} 
\end{figure}

\begin{figure}
 \epsscale{1.0}
   \plotone{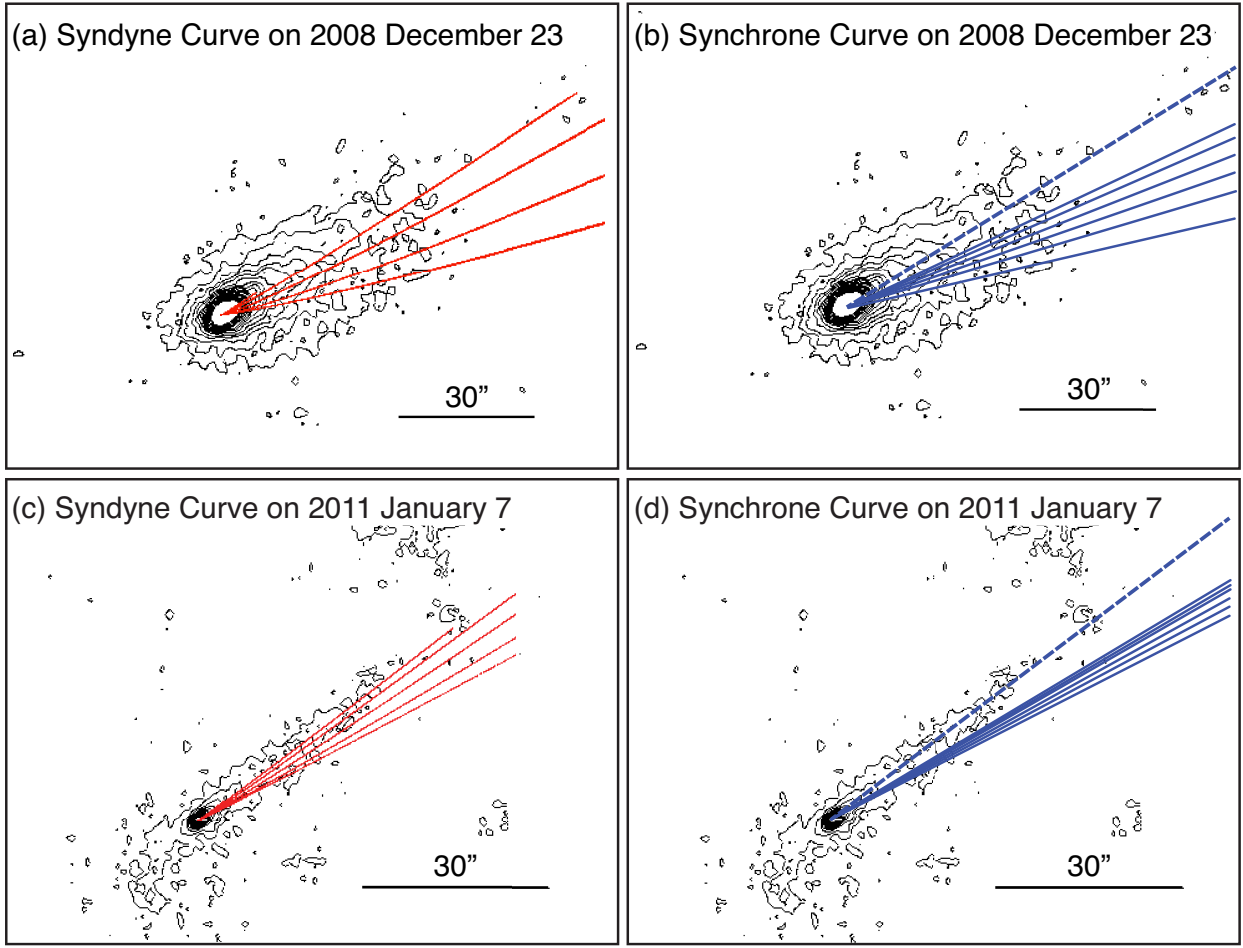}
  \caption{Syndyne and synchrone curves of 17P/Holmes on 23
 December 2008 and 7 January 2011.  The syndynes curves are
 characterized  by $\beta_{rp}=0.57$ (1 \micron), 0.057  (10 \micron), 
 0.0057 (100 \micron), 0.00057 (1 mm), and 0.000057 (1 cm), rotating
 in anticlockwise direction. The synchrone curves are  characterized by  
 the ejection times 30, 60, 90, 120, 150, 180 days before the observed
 day from bottom up. The top dashed lines are the synchrone curves
 ejected on 2007 October 23 when the outburst occurred.
}\label{fig:sync-synd}  
\end{figure}

In the second paragraph of Section 3.1, we provided a simple theory to
explain the cause of broadening in Figure \ref{fig:radi}, which helps us
to understand the broadening in a straightforward manner. A similar
theory was given in \citet{Stevenson2012}. It is important to check
the consistency of the discussion with the three dimensional model we
provided in Section 3.5. We examined whether the outburst dust particles
contributed to the broadening in our images using 
our three dimensional model. The motion of the dust particles is
more complicated than we discussed in Section 3.2. Dust particles
expand because of the initial velocity, but at the same time, some dust
particles ejected sunward are  affected by the solar radiation
pressure. Figure \ref{fig:appendix} shows 
time-series simulation images of dust particles ejected
on 23 October 2007 based on the three dimensional model. There is no
circular envelope as observed in Figure 2 (a) and (b) most likely because
the model simulates the low velocity component. In the figure, two
circles denote the aperture radius of 0.2\arcsec~ and 1\arcsec, 
respectively. Within the aperture, we observed the broadening in all
data sets. We found that the brightest part of the cloud, which could
contribute the broadening of radial profiles, was detached $\sim$10
months after the outburst. Since our postburst data were taken $>$11
months after the outburst, dust particles ejected by the outburst could
not contribute the broadening of the radial profiles. It is natural to
think 17P/Holmes showed the broadened brightness profile because it was
active when we made the observations.


\clearpage


\clearpage


\clearpage


\clearpage



\clearpage


\clearpage


\clearpage


\clearpage


\clearpage


\clearpage


\begin{figure}
 \epsscale{0.6}
   \plotone{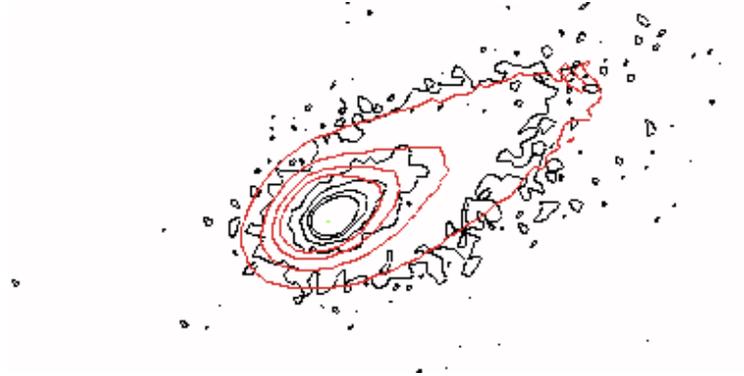}
  \caption{Comparison between observed and modeled contour maps on 23   December 2008. }\label{fig:model} 
\end{figure}

\clearpage


\begin{figure}
 \epsscale{0.8}
   \plotone{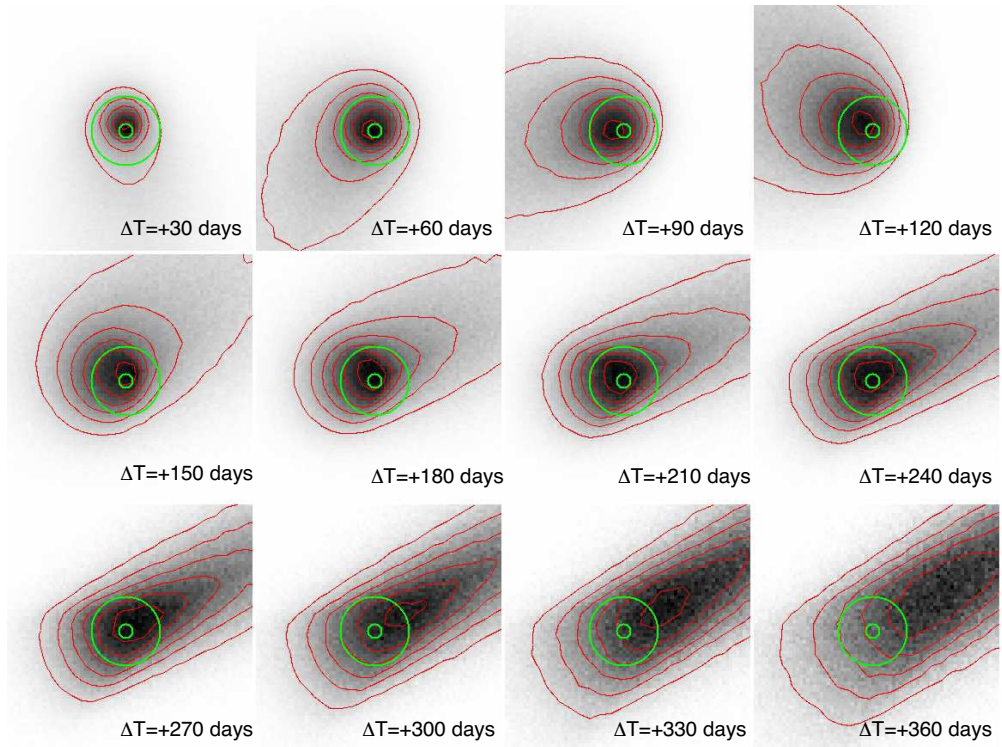}
  \caption{Motion of dust particles ejected by 2007 outburst based on a
 model in Section 3.5. Each panel denotes the simulation image with
 contour map on a day $\Delta$T after the outburst. Two circles
 correspond apparent angular  distances of  0.2\arcsec~ and
 1\arcsec. These images are the standard  orientation in  the sky, that
 is, Celestial North is up and East is to  the left. The  field of view
 is $7\arcsec\times 7\arcsec$.} \label{fig:appendix}
\end{figure}

\clearpage


\begin{table}
  \caption{Observational circumstance. $r_h$, $\Delta$, and $\alpha$
 denote, respectively, the median heliocentric  distance (AU), the median geocentric
 distance (AU), and the median Sun--Comet--Observer angle (\arcdeg).
 UT and Exptime indicate the median observation date (UT) and total
 exposure time in minutes. Position data  are obtained  from our
 numerical calculation by using the  orbital elements in JPL's  on-line
 site.}  
 \label{tab:cirsumstance}
  \begin{center}
    \begin{tabular}{lrrrrrrrr}
     \hline
Observatory & Instrument & UT & Filter & Exptime & $r_h$ &
     $\Delta$ & $\alpha$ & $\Delta T$\\   
 \hline

\\(Perihelion) &    & 2007/05/05.04 &                &      & 2.05 &     &      & $-$171.3\\
\\
AKARI  & MIR        & 2007/08/23.36 & L18W            & --  & 2.23 & 1.99 & 27.0 & $-$60.9\\
\\
(Outburst)   &      & 2007/10/23.30 &                &      & 2.43 &      &     & 0.0\\
\\
Subaru & Suprime-Cam& 2008/09/29.62 & VR & 4.5  & 3.84 & 4.26 &12.9 & +342.3\\
NHAO & MINT       & 2008/12/23.71 & $R_\mathrm{C}$ & 297  & 4.13 & 3.35 & 9.2 & +427.4\\
NHAO & MINT       & 2008/12/27.70 & $R_\mathrm{C}$ &  258  & 4.15 & 3.32 & 8.3 & +431.4\\
NHAO & MINT       & 2010/01/16.86 & $R_\mathrm{C}$ &  33  & 5.01 & 4.27 & 8.0 & +816.6\\
\\
(Aphelion)   &      & 2010/10/14.25 &                &      & 5.19 &      &     &\\
\\
Subaru & Suprime-Cam& 2011/01/07.57 & $R_\mathrm{C}$ &  30  & 5.17 & 4.90 &10.8 & +1172.3\\
UH2.2m  & Tex2k      & 2011/02/04.48 & $R_\mathrm{C}$ & 114  & 5.16 & 4.48 & 8.6 & +1200.2\\
UH2.2m  & Tex2k      & 2011/02/05.53 & $R_\mathrm{C}$ & 216  & 5.16 & 4.47 & 8.5 & +1201.2\\
HCT    & Site2k$\times$4k   & 2011/03/29.85 & $R_\mathrm{C}$ &  87  &
                         5.12 & 4.16 & 3.2 & +1253.6\\
HCT    & Site2k$\times$4k   & 2011/03/30.76 & Free           & 231  &
                         5.12 & 4.16 & 3.7 & +1254.5\\
Subaru & Suprime-Cam& 2011/06/06.28 & g$'$, r$'$, z$'$     & 4.5, 3, 15 &
                         5.06 & 4.82 & 11.5 & +1322.0\\
\\
(Perihelion) &      & 2014/03/03.16 &                &      & 2.05 &
                             &     & +2345.1\\

\\  \hline
    \end{tabular}
  \end{center}
\end{table}

\begin{table}
  \caption{Photometric results.}
  \begin{center}
    \begin{tabular}{lrrrrrrr}
\hline
UT & $m_R$ [error] & $m_R(1,1,0)$ & $\phi$ [\arcsec] & $\eta$ & $\dot{M_d}$ [kg s$^{-1}$] \\
\hline
2005/03/06.00 & 22.86 [0.15] & 16.24 & 1.00 &$<$0.14 &$<$0.17 \\
2007/08/23.36 &  0.39$^\dag$ [0.04] &   --  & 21.0 &   17.0 &   2.75 \\
2007/10/27.65 &  2.46 [0.20] & $-$1.11 &   -- &    --  &  --    \\
2008/12/23.71 & 18.99 [0.19] & 12.96 & 4.75 &   17.3 &   5.40 \\
2008/12/26.81 & 18.82 [0.15] & 12.83 & 5.00 &   19.8 &   5.89 \\
2010/01/16.86 & 21.81 [0.22] & 14.97 & 4.00 &   2.13 &   0.56 \\
2011/01/07.57 & 22.60 [0.15] & 15.20 & 1.75 &   1.32 &   0.68 \\
2011/02/05.53 & 22.08 [0.12] & 14.96 & 2.50 &   1.91 &   0.76 \\
2011/03/29.85 & 21.22 [0.26] & 14.47 & 6.50 &   3.58 &   0.59 \\
\hline
    \end{tabular}
  \end{center}
 \label{tab:photometry}
\tablecomments{
$^\dag$ Infrared flux [Jy] at 18 \micron.\\
}
\end{table}

\begin{table}
  \caption{Input and best-fit parameters for the dust ejection model.}
  \begin{center}
    \begin{tabular}{llc}
\hline
Parameter   & Input values & Best-fit values \\
\hline
$u_1$ & 0.1--0.5 with 0.1 interval & 0.3--0.5 \\
$u_2$ & 0.1--0.5 with 0.1 interval & 0.2--0.4 \\
$q$ & 3.0--4.0 with 0.1 interval & 3.4--3.6 \\
$k$ & 0--12 with 3 interval & 9\\
$a_\mathrm{min}$ [\micron] & 1 (fixed) & 1\\
$a_\mathrm{max}$ [\micron] & 10$^2$, 10$^3$, 10$^4$  & 10$^3$--10$^4$ \\
$V_0$ [m/s] & 10--150 with 10 interval & 30--35 \\
$\omega$ [\arcdeg] & 10--60 with 5 interval & 15--25\\

\hline
    \end{tabular}
  \end{center}
 \label{tab:parameter}
\end{table}

\end{document}